\documentclass[10pt]{article}

\usepackage[margin=1in]{geometry}
\usepackage[backend=biber, doi=true, natbib=true]{biblatex}
\usepackage{authblk}
\usepackage{graphicx}
\usepackage{xltabular}
\usepackage{booktabs, caption, longtable, colortbl, array}
\usepackage{threeparttablex}
\usepackage{pdflscape}

\begin{document}

\title{Assessing Acceptance and Privacy Preferences of Third-party Financial Data Sharing in Bipolar Disorder}

\author[1]{Jeff Brozena}

\author[1]{Johnna Blair}

\author[2]{Dahlia Mukherjee}

\author[2]{Erika F. H. Saunders}

\author[3]{Thomas Richardson}

\author[4]{Mark Matthews}

\author[1]{Saeed Abdullah}

\affil[1]{College of Information Sciences and Technology, Penn State University, Pennsylvania, USA}

\affil[2]{Penn State College of Medicine, Pennsylvania, USA}

\affil[3]{School of Psychology, University of Southampton, UK}

\affil[4]{School of Computer Science, University College Dublin, Ireland}

\maketitle

\abstract{

\textbf{Objectives} Bipolar disorder is strongly associated with financial instability. We examine how different interventions motivate individuals with bipolar disorder to share financial data with others. This approach can inform the development of tools for digital monitoring and intervention designed to promote financial stability in this population.

\textbf{Methods} 500 individuals with BD completed a pre-registered factorial vignette survey to examine level of comfort with hypothetical scenarios involving third-party financial interventions during symptomatic and euthymic periods. Scenario components were systematically varied between third-party actors, mood states, and intervention types. Participants rated sharing comfort on a 0-10 point scale. Multilevel models tested differences alongside clinical and financial histories, relational trust, and personality.

\textbf{Results} Participants were most comfortable involving care partners in financial planning. They were more comfortable with temporary spending restrictions during symptomatic states than euthymic periods, underscoring the importance of accurate mood detection for intervention delivery. Prior financial help-seeking behavior and higher relational trust predicted greater comfort. Bankruptcy experience --- declared by 11.4\% and considered by 31.7\% --- was associated with increased comfort with spending restrictions. Individuals with psychiatric advance directives (8\%) were significantly more comfortable sharing spending behaviors than those without.

\textbf{Conclusions} Comfort with financial interventions was higher among those with prior financial challenges or help-seeking histories. Participants distinguished between symptomatic and euthymic periods, favoring targeted, time-limited restrictions over general monitoring. These findings extend prior work on financial data sharing for illness self-management, highlighting the role of trusted third parties in designing acceptable, effective interventions.
}

\section{Introduction}

Individuals living with bipolar disorder (BD) are at an increased risk of financial instability \cite{richardsonfinancial2018}. Risky or impulsive decision-making can lead to serious long-term financial instability, which can severely impact the quality of life for individuals with BD and their care partners. For example, 70\% of individuals with BD have reported impulsive spending during hypomania \cite{fletcherhighrisk2013}. A population-scale study of consumer credit histories and health records showed that individuals with BD-I are 50\% more likely to have declared bankruptcy in their lifetime than the healthy population \cite{nauassessment2023}. The complex interplay of finances and bipolar disorder has been shown to be cyclic and mutually reinforcing, with mood symptoms leading to compulsive buying leading to worsening mood symptoms \cite{richardsonfinancial2018}. The impact of financial instability often extends beyond the individual: Financial difficulties have been cited by caregivers as among the most burdensome effects of mania \cite{beentjescaregiver2012}.

The frequency and severity of illness-related financial issues warrant novel approaches tailored to support long-term financial stability in bipolar disorder. Towards this end, digital health applications are appropriate given their capacity to situate into the daily lives of individuals and into existing networks of care. Advances in open banking technologies have made financial data available to contexts other than banking applications. Financial data can be repurposed to aid in tasks like financial planning for illness-related crises, monitoring for early signs of relapse, and (if desired) invoking pre-arranged interventions during periods of heightened financial risk. Supportive financial technologies for bipolar disorder can offer more than guidance to individuals and trusted third parties. For example, financial technologies can be designed to contain conceptual similarities with informational psychiatric advance directives such that they respect the rights, personal preferences, and privacy of an individual throughout the course of their illness.

Prior work supports the acceptance of this direction by individuals with bipolar disorder. Our research has identified that individuals with BD are, on average, quite comfortable incorporating their financial data into supportive self-management tools \cite{brozenasupportive2024}. Separately, individuals living with serious mental illness have demonstrated they are comfortable sharing financial data to notify care partners of potential financial issues \cite{barrospenafinancial2021}. These findings collectively signal the acceptance of incorporating financial data into illness management. However, there is limited research investigating acceptance of involving trusted care partners in ongoing financial management in illness-related contexts.

To address this gap, we aimed to survey individuals living with bipolar disorder to assess level of comfort with financial data sharing across a variety of contexts. We specifically focused on understanding whether individuals with BD were comfortable sharing financial data with others who could potentially offer aid, accountability, or early intervention into risky financial behaviors.

\section{Materials and methods}

\subsection{Preregistered Research Design}

We conducted a survey experiment (N=500) to assess level of comfort of individuals with bipolar disorder with financial data sharing for financial interventions involving different stakeholders. Our study design, hypotheses, and analysis plan were preregistered with the Open Science Foundation \cite{foster2017open}. We have made these materials and analysis scripts available in a public repository \cite{brozenaidentifying2023}.

Participants were recruited using Prolific in February, 2024 from a representative sample of the U.S. population. Individuals were eligible to participate if they self-reported a primary diagnosis of bipolar disorder (of any type), were 18 years or older, lived in the U.S., and spoke English. Individuals were excluded if they could not provide consent, did not speak English, were under the age of 18, or did not report a primary diagnosis of BD.

We chose to implement a factorial vignette design. This approach allows researchers to assess the relative importance of dimensions (``factors'') under analysis by experimentally varying aspects of those dimensions (``levels'') within a structured vignette template \cite{auspurgfactorial2015}. Participants were presented with all vignette possibilities and asked to rate their level of comfort with each resulting scenario on a scale of 0 -- 10, with 0 being completely uncomfortable and 10 being completely comfortable. We randomized the vignette display order to control for context effects. See Table \ref{tbl-factors} below for details on the template we implemented.

Ethics approval was granted by Pennsylvania State University (Reference 00019759) and by University of Southampton (Reference 74392). 

\subsubsection{Vignette Dimensions}

We selected factors to explore how third parties might aid and intervene in the financial lives of individuals with BD. The vignette experiment contained three factors: \emph{actor} (care partners, clinicians, banks), \emph{intervention context} (monitoring spending details, help with planning and budgeting, 48-hour spending restriction), and \emph{mood state} (symptomatic or euthymic). This resulted in 18 unique vignettes (3 x 3 x 2).

\begin{table}[ht]
    \centering
    \caption{Participants were asked to rate their level of comfort with 18 total scenarios containing random combinations of these factors and their levels. \textit{Structure of vignettes:} "The app uses your financial data so that it can [Intervention Context] with your [Actor] [Mood State]."}
    \label{tbl-factors}
    \begin{tabular}{l l}
        \hline
        \textbf{Factor} & \textbf{Levels} \\
        \hline
        Actor & 
        \begin{tabular}[t]{@{}l@{}}
            Care Partner \\
            Clinician \\
            Banks
        \end{tabular} \\
        \hline
        Intervention Context & 
        \begin{tabular}[t]{@{}l@{}}
            Share spending details \\
            Planning \& budgeting \\
            48-hour spending restriction
        \end{tabular} \\
        \hline
        Mood State & 
        \begin{tabular}[t]{@{}l@{}}
            During a mood episode \\
            During stable mood
        \end{tabular} \\
        \hline
    \end{tabular}
\end{table}

\paragraph{Mood State}

We primarily sought to assess any differences between the presence or absence of symptoms. As such, we chose not to discern between mania (or hypomania) and depression, opting for two levels representing the presence or absence of symptoms of any type. Additionally, given the high degree of variability between individuals, we did not want to assume that risky or otherwise problematic financial behaviors occurred \emph{only} during \emph{specific} illness states \cite{richardsonfinancial2018}.

We hypothesized that respondents' level of comfort with data sharing would show a significant overall decrease during symptomatic periods when compared to euthymic periods (H1.1), reflecting both an awareness of illness-related vulnerability along with a desire to preserve autonomy during euthymic periods. Additionally, we hypothesized that \emph{sharing spending details} during symptomatic periods would result in the highest level of comfort across combinations of the factors Mood State and Intervention Context (H1.2). Finally, we hypothesized that comfort ratings for 48-hour spending restrictions during symptomatic periods would be higher than during euthymic periods (H1.3).

\paragraph{Actor}

We included third parties who could, hypothetically, play a role in financial data sharing and intervention. \emph{Care partners} were suggested to represent a spouse or other family member, friends, or a formal or informal caregiver. We chose not to differentiate between clinical role or license, opting for a \emph{clinician} level to represent these relationships in aggregate. Finally, although financial data may be readily available via other channels, we chose to include \emph{banks} given their immediate existing access to this data and their high capacity to implement a given intervention. We hypothesized that respondents would be most comfortable involving care partners when compared with other actors (H1.3).

\paragraph{Intervention Context}

We sought to understand the complex interplay between perceived vulnerability to illness-specific financial risks and dynamics of agency, privacy, and autonomy. As such, we chose to include a set of hypothetical third-party interventions ranging in severity, level of data access, and involvement. \emph{Sharing spending details} would, ideally, involve others in the passive monitoring of financial behaviors. \emph{Help with planning and budgeting} would provide collaborative interactions designed to involve others in tasks related to meeting financial goals. A hypothetical \emph{48-hour spending restriction} would allow for a ``cooling-off period'' \cite{blairfinancial2022, farrwhy2019} if third parties believed an individual were exhibiting risky financial behaviors. We hypothesized that respondents would be least comfortable with 48-hour spending restrictions when compared to other intervention contexts (H2.1).

\subsubsection{Other Measures}

We collected self-reported demographic measures of race, ethnicity, education level, age, gender, marital status, and income. We included additional measures of respondents' prior experiences related to clinical and financial histories that were necessary to test our hypotheses and to conduct exploratory analyses. In the sections that follow, we detail our rationale for including these measures.

\paragraph{Clinical History}

We sought to analyze differences in comfort based on details related to diagnosis. To that end, we collected respondents' diagnostic subtype, providing the option to select ``I do not know''.

We hypothesized that level of comfort with data sharing would increase given the presence of advance care planning (H2.2). We operationalized evidence of advance care planning as having a \emph{psychiatric advance directive}, hypothesizing that respondents who had implemented one would be more comfortable sharing spending details than those who had not (H2.2). In both instances, respondents selected ``Yes'', ``I do not know'', or ``No''.

Given the increased risk of having declared bankruptcy in this population \cite{nauassessment2023}, we prompted respondents whether they had (1) considered bankruptcy or (2) previously declared it. We hypothesized that those who had declared bankruptcy, keenly aware of financial consequences of this illness, would be more comfortable with a 48-hour spending restriction than those who had not (H2.3).

\paragraph{Financial History}

We included variables associated with respondents' primary financial goals, help-seeking behaviors specific to financial impulsivity, and evidence of prior adverse financial events. We asked respondents to identify their financial goal. We provided them a list of financial goals based on prior survey findings.

Our prior work indicates that increased willingness to involve others for help around money issues is associated with higher comfort in third-party financial data sharing \cite{brozenasupportive2024}. In order to better assess how help-seeking behavior might predict increased comfort in this context, we asked respondents specifically whether they had ever asked for help managing impulsive spending, hypothesizing that those answering ``Yes'' would predict higher comfort ratings for all third party actors (H3.3).

We asked respondents to rate how easy or difficult it was to trust each respective actor on a 5-point Likert scale. For each actor, we compared those who reported it was ``extremely easy'' to those who reported it was ``extremely difficult'' to trust the actor.

As an exploratory measure of debt-related behaviors, we asked participants whether they had ever used a Buy Now, Pay Later service.

\paragraph{Validated Scales}

We included a standard BFI-10 scale \cite{rammstedtmeasuring2007}, hypothesizing that agreeableness would have a significant, non-directional effect on level of comfort (H4.1).

\subsection{Statistical Methods}

We used multilevel linear models \cite{gelmanData2007} to account for the vignette experiment’s hierarchical data structure (i.e., vignettes within respondents). We first identified a primary model incorporating variables of interest, then created nested models contained within the primary model. In all models, participants and vignette items were selected as random effects. We calculated the intraclass correlation coefficient (ICC) to account for similarity between groups in these random effects. In this case, a higher ICC value indicates greater similarity between groups (e.g., between participants) for a given random effect. Finally, we performed a type-3 ANOVA on the primary model to identify significant effects. All modeling was performed in R 4.5.0 using {lme4} \cite{bateslme42025}. 

Post-estimation processing was performed using {marginaleffects} \cite{arel-bundockhow2024} in R. During this processing we calculate the expected value of the outcome variable for different subgroups of the sample. Expected values are first computed as predicted values at each row of the dataset given specific predictors of interest, then by averaging these predicted values across subgroups of interest. In cases where we are testing a hypothesis involving differences between subgroups, we first compute the average predictions of interest, then specify that hypothesis within the {marginaleffects} function call. In all cases, we report these quantities as estimates with standard errors. In the case of a hypothesis test, we also report a \textit{p}-value.

Following our preregistration \cite{brozenaidentifying2023}, we tested hypotheses regarding mood state (H1.1, H1.2, H1.3), intervention context (H2.1, H2.2, H2.3), and actors (H3.1, H3.2, H3.3). Hypotheses H1.3, H2.1, and H3.1 are primarily concerned with the main effects of the vignette factors, while H1.1, H1.2, and H3.2 are concerned with their interaction effects. H2.2, H2.3, H2.4, and H4.1 are concerned with the interaction effects of vignette factors with respondent characteristics.

\section{Results}

\subsection{Demographics and Data Quality}

The mean rating across all vignettes for 500 respondents was 4.18 (SD=3.26) on a scale of 0--10, with 0 representing complete discomfort and 10 representing complete comfort. Table 2 shows the descriptive results and 95\% confidence intervals for each of the 18 vignettes.

\begin{table}[ht]
\centering
\caption{Descriptive statistics for each of the 18 vignette scenarios. 
Participants rated their level of comfort on a scale from 0–10, with 0 representing 
complete lack of comfort and 10 representing complete comfort.}
\resizebox{.85\textwidth}{!}{
\begin{tabular}{rlllrrrrrr}
  \hline
 & \textbf{Actors} & \textbf{Intervention Context} & \textbf{Mood State} & \textbf{Mean} & \textbf{Median} & \textbf{SD} & \textbf{Lower} & \textbf{Upper} & \textbf{Obs} \\ 
  \hline
1 & Banks & 48-hour Spending Restriction & Euthymic & 1.81 & 0.00 & 2.54 & 1.59 & 2.03 & 500 \\ 
  2 & Banks & 48-hour Spending Restriction & Symptomatic & 2.74 & 2.00 & 2.96 & 2.48 & 3.00 & 500 \\ 
  3 & Banks & Help with Planning \& Budgeting & Euthymic & 4.26 & 4.00 & 3.18 & 3.98 & 4.54 & 500 \\ 
  4 & Banks & Help with Planning \& Budgeting & Symptomatic & 3.81 & 4.00 & 3.20 & 3.52 & 4.09 & 500 \\ 
  5 & Banks & Share Spending Details & Euthymic & 3.16 & 2.00 & 3.03 & 2.89 & 3.43 & 500 \\ 
  6 & Banks & Share Spending Details & Symptomatic & 2.88 & 2.00 & 2.96 & 2.62 & 3.14 & 500 \\ 
  7 & Care Partner & 48-hour Spending Restriction & Euthymic & 2.98 & 2.00 & 2.98 & 2.72 & 3.24 & 500 \\ 
  8 & Care Partner & 48-hour Spending Restriction & Symptomatic & 4.58 & 5.00 & 3.22 & 4.30 & 4.86 & 500 \\ 
  9 & Care Partner & Help with Planning \& Budgeting & Euthymic & 5.94 & 6.00 & 2.94 & 5.68 & 6.20 & 500 \\ 
  10 & Care Partner & Help with Planning \& Budgeting & Symptomatic & 5.95 & 6.00 & 2.96 & 5.69 & 6.21 & 500 \\ 
  11 & Care Partner & Share Spending Details & Euthymic & 5.06 & 5.00 & 3.12 & 4.78 & 5.33 & 500 \\ 
  12 & Care Partner & Share Spending Details & Symptomatic & 5.41 & 6.00 & 3.06 & 5.14 & 5.68 & 500 \\ 
  13 & Clinicians & 48-hour Spending Restriction & Euthymic & 2.58 & 2.00 & 2.81 & 2.33 & 2.82 & 500 \\ 
  14 & Clinicians & 48-hour Spending Restriction & Symptomatic & 3.74 & 4.00 & 3.10 & 3.46 & 4.01 & 500 \\ 
  15 & Clinicians & Help with Planning \& Budgeting & Euthymic & 5.29 & 6.00 & 3.03 & 5.03 & 5.56 & 500 \\ 
  16 & Clinicians & Help with Planning \& Budgeting & Symptomatic & 5.40 & 6.00 & 3.00 & 5.13 & 5.66 & 500 \\ 
  17 & Clinicians & Share Spending Details & Euthymic & 4.63 & 5.00 & 3.08 & 4.36 & 4.90 & 500 \\ 
  18 & Clinicians & Share Spending Details & Symptomatic & 5.03 & 6.00 & 3.12 & 4.75 & 5.30 & 500 \\ 
   \hline
\end{tabular}}
\end{table}

Our respondents were mostly female (59.9\%), aged 35 - 44 (24.8\%), university-educated (30.1\%), and employed full-time (41.4\%). BD-II was the most common diagnosis (43.3\%), with 23\% reporting a BD-1 diagnosis and 23.8\% reporting BD Not Otherwise Specified. The majority received their BD diagnosis when aged 19 to 29 years. \textbf{11.4\% of respondents had declared bankruptcy and 31.7\% had considered it as a possibility.} A complete descriptive table of demographics and other measures of our sample are available in Appendix \ref{app-1}.

The median response time to the survey was 10 minutes. Chi-squared tests yielded no systematic demographic dependencies, confirming the randomized experimental assignment of the vignettes worked as intended. As such, we did not control for demographic variables in our submodels. Demographic variables were included in our full model. Additionally, estimated marginal predictions at each variable level under analysis is available in Appendix \ref{app-2}.

The intraclass correlation coefficients (ICC) were 0.398 and 0.148 for Participant ID and Vignette Item respectively. In other words, 39.80\% of the variance is due to between-group differences in participants, where 14.80\% is due to between-group differences in vignette scenarios. A higher ICC value suggests greater similarity between groups. Given the lower explained variance between groups, these results indicate the necessity of multilevel modeling for this sample.

\subsection{Preregistered Hypotheses}

\subsubsection{Differences in intervention comfort across the illness trajectory}

We hypothesized that respondents would be more comfortable with all intervention contexts during periods of mood symptoms versus periods of euthymia (H1.1). As no significant differences in comfort rating were observed in a comparison between these two periods, we reject H1.1 (estimate=0.42, SE=0.21, \emph{p}=0.04).

\begin{figure}[!ht]
  \centering
  \includegraphics[scale=0.6]{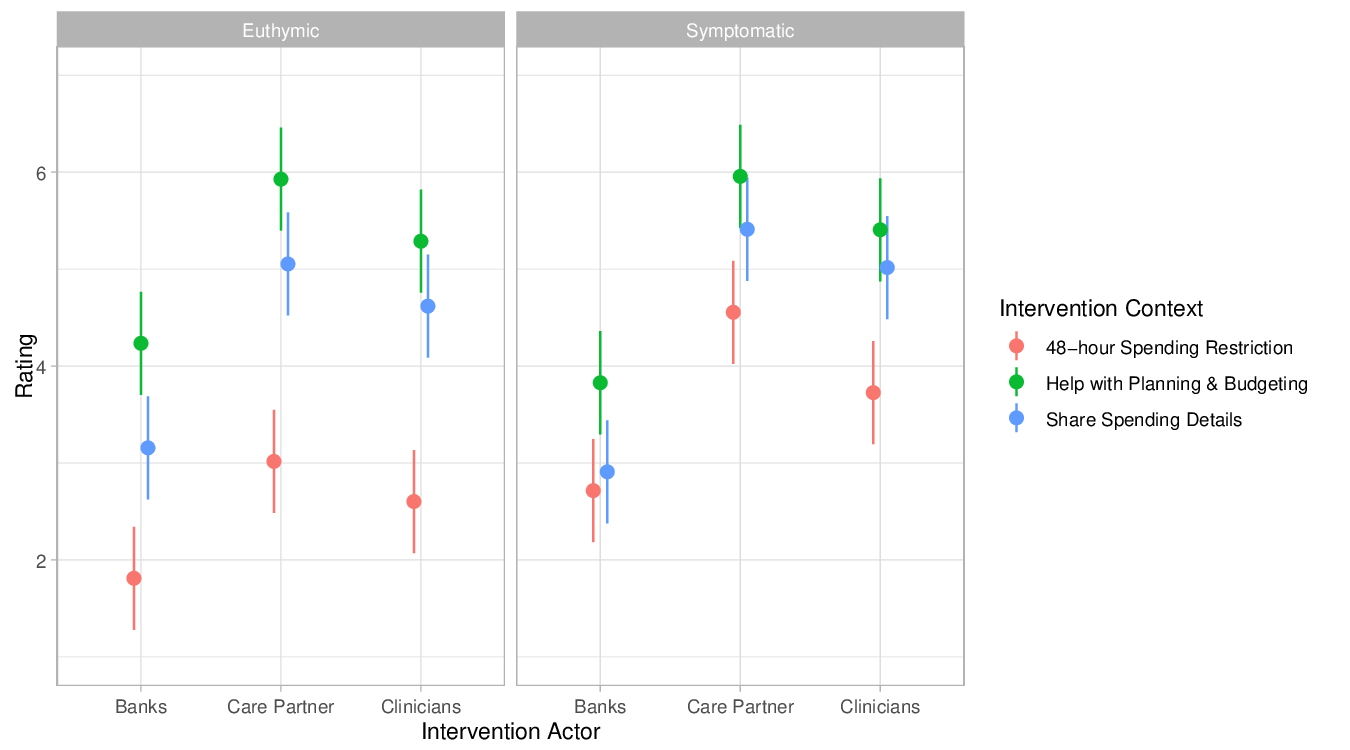}
  \caption{Marginal predictions of comfort levels averaged across intervention contexts and actors on a scale of 0 -- 10. Respondents were most comfortable seeking third-party help with planning and budgeting, especially from their care partners.}
  \label{fig:expl_main_actors_context_moodstate}
\end{figure}

We hypothesized that participants would be most comfortable sharing spending behaviors during periods of mood symptoms (H1.2). We expected respondents to be most comfortable with passive interventions (e.g., that spending details could be shared automatically) during symptomatic periods. However, respondents in our sample were most comfortable seeking help with planning and budgeting during both symptomatic and euthymic periods. As such, we reject hypothesis H1.2. 

Finally, we wanted to understand how respondents viewed hypothetical third-party spending restrictions. Given its potential for impact on individual autonomy and decision-making, we hypothesized that temporary spending restrictions of any sort would receive the lowest comfort ratings among intervention types (H2.1). While this hypothesis was ultimately supported (estimate=-1.11, SE=0.15, \emph{p}\textless0.01), several factors were associated with increased spending restriction comfort, including whether respondents had experienced a prior bankruptcy or hospitalization. We will discuss those factors in greater depth below.

\subsubsection{Financial help-seeking behavior and third-party trust}

In this section, we explore several relational factors that may precede comfort with involving third parties in financial interventions. First, we hypothesized that respondents would be most comfortable involving their care partners (rather than their clinicians or banks) in financial interventions (H3.1), a claim supported in our analysis (estimate=4.99, SE=0.02).

We hypothesized that higher levels of trust in each intervention actor would be associated with a greater level of comfort with the involvement of each respective actor in an intervention (H3.2). For all actors, respondents who reported it was ``extremely easy'' were significantly more comfortable involving them in an intervention than those who reported low levels of trust (i.e., those who reported it was "extremely difficult" to trust respective third parties). Marginal predictions for these variables are available in Appendix \ref{app-2}.

\begin{figure}[!ht]
  \centering
  \includegraphics[scale=0.5]{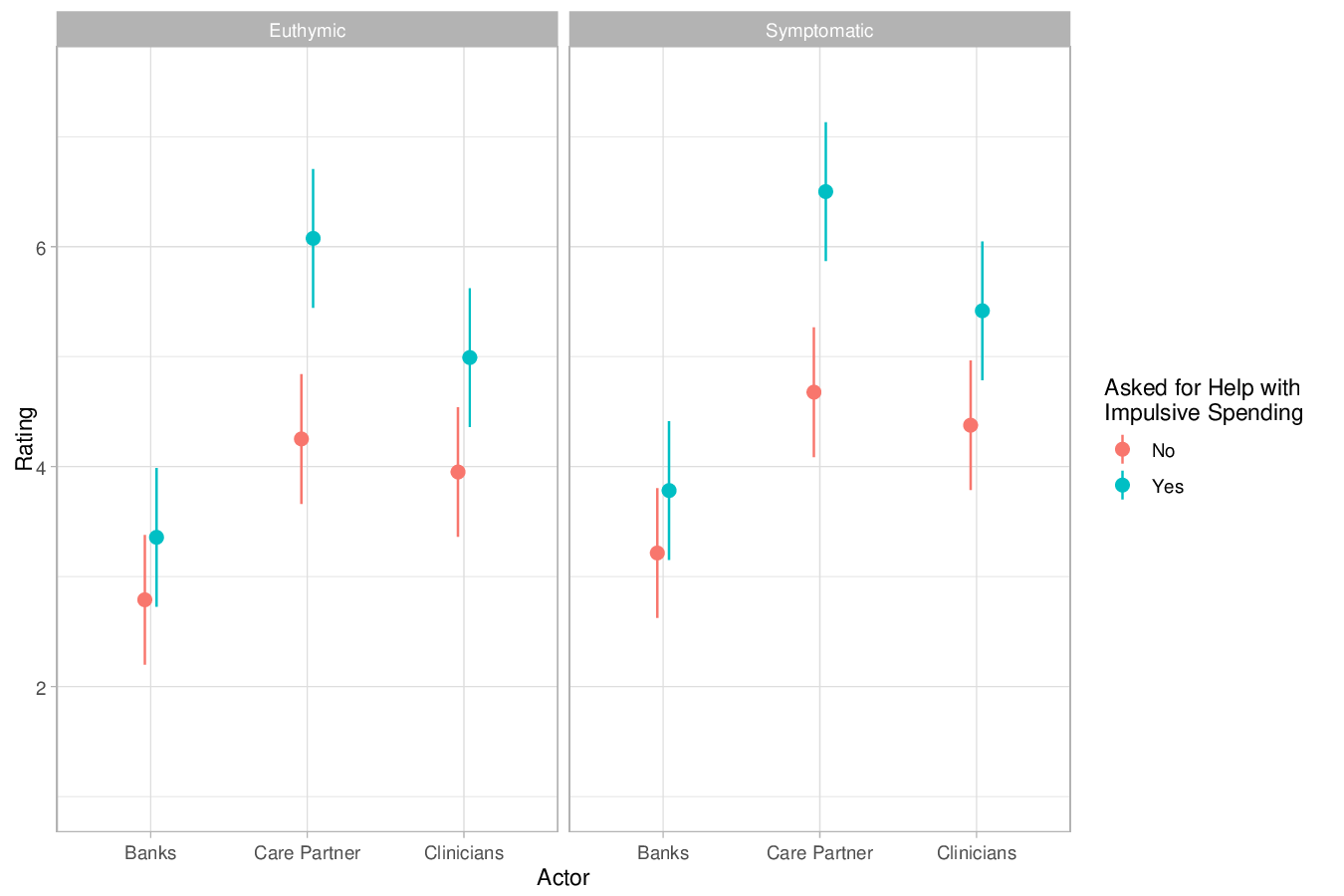}
  \caption{Prior financial help-seeking behaviors were associated with a significant increase in comfort ratings across all third-party intervention actors.}
  \label{fig:expl_interaction_carepartnerhelp_actors_moodstate}
\end{figure}

Next, we specifically considered the influence of prior help-seeking behavior on level of comfort. We hypothesized that respondents who had previously asked their care partners for help with impulsive spending would be significantly more comfortable involving all actors (H3.3). This hypothesis was supported, as a significant difference in comfort ratings was observed in a comparison between those who had asked for help to those who had not (estimate=1.14, SE=0.19, \emph{p}\textless0.001). Figure \ref{fig:expl_interaction_carepartnerhelp_actors_moodstate} displays these differences as marginal predictions for each intervention actor. 

\subsubsection{Advanced planning and financial history as a predictor of intervention comfort}

Next, we wanted to understand how certain illness-related lived experiences might influence intervention comfort. We included factors associated with the preservation of future autonomy (i.e., the creation of a psychiatric advance directive) and the consequences of past financial behavior (i.e., prior bankruptcy).

We questioned whether those who take an active role in planning for future episodes might be more comfortable sharing spending information with third parties (H2.2). As these individuals had made the effort to legally formalize their care planning wishes, we expected a heightened level of openness to passive interventions that could support their future care plans -- here, the passive sharing of spending details. In a comparison between those with and without psychiatric advance directives, we observed a significant increase in level of comfort with sharing their spending information (estimate=1.08, SE=0.37, \emph{p}=0.003), supporting H2.2.

We hypothesized that those who had previously experienced adverse financial events, operationalized as prior bankruptcy, would be more comfortable with spending restrictions given their lived experience of these financial consequences (H2.3). Respondents who had declared bankruptcy reported significantly higher levels of comfort with spending restrictions than those who had not (estimate=0.82, SE=0.32, \emph{p}=0.01), supporting H2.3.

\begin{figure}[!ht]
  \centering
  \includegraphics[scale=0.5]{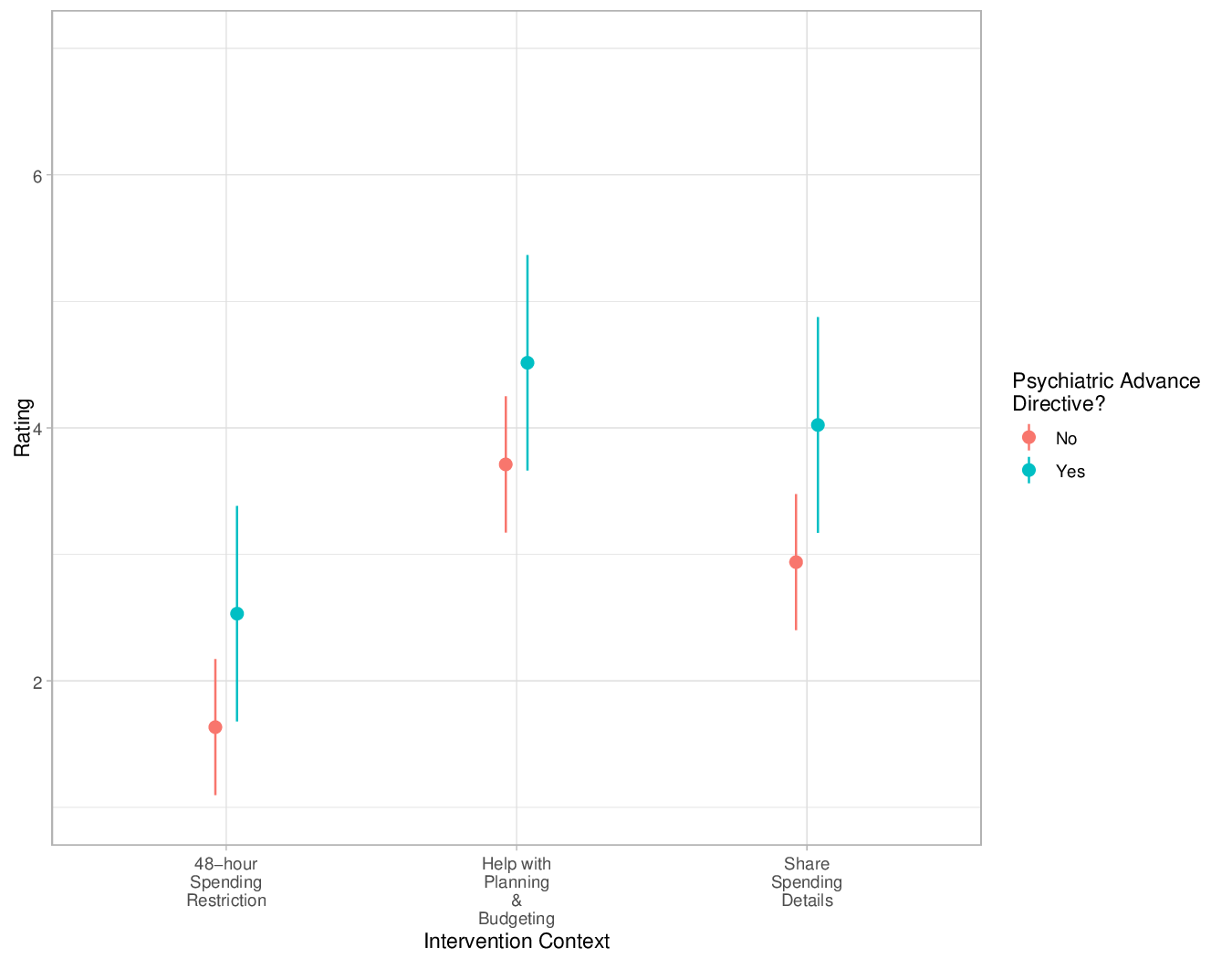}
  \caption{Advance planning behaviors (i.e., psychiatric advance directives) were associated with greater comfort with all hypothetical interventions.}
  \label{fig:H2.2_interaction_context_PAD}
\end{figure}

\subsubsection{The influence of personality on intervention comfort}

Given the relationship between concern for privacy and agreeableness \cite{najafianfactors2021, junglaspersonality2008}, we hypothesized that agreeableness would have a significant non-directional effect on level of comfort (H4.1). This finding was supported in a type III ANOVA (Chisq=8.75, \emph{p}=0.003), leading us to accept H4.1. We also explored the association between other personality dimensions and level of comfort with data sharing in the following section.

\subsection{Exploratory Analysis}\label{exploratory-analysis}

\begin{figure}[!ht]
  \centering
  \includegraphics[scale=0.6]{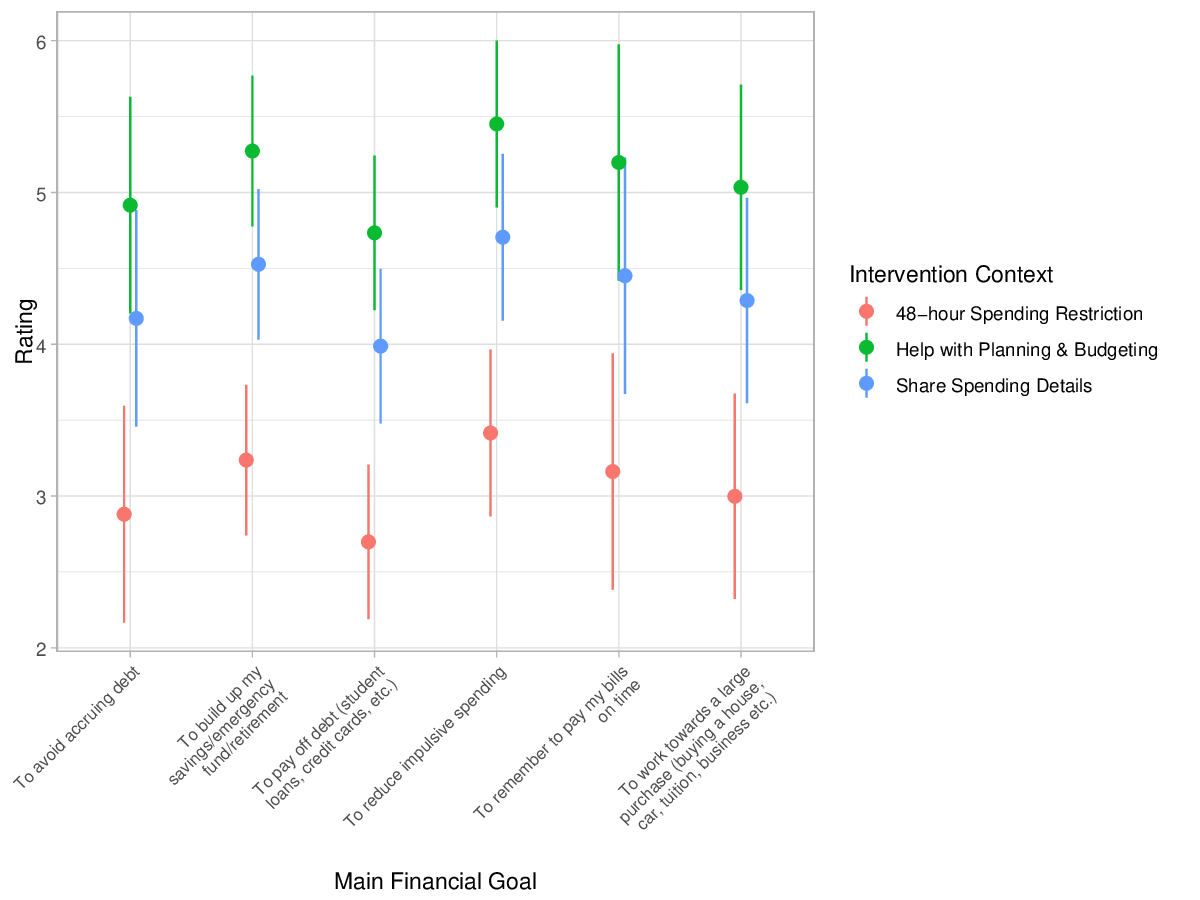}
  \caption{Individuals whose main financial goal was to reduce impulsive spending were most comfortable with all hypothetical intervention contexts.}
  \label{fig:expl_main_financialgoal_context}
\end{figure}

In an unregistered descriptive analysis of survey data, we explored the interplay of primary financial goals and level of comfort with financial data sharing for specific interventions. Individuals whose primary financial goal was ``to reduce impulsive spending'' were most comfortable with all hypothetical interventions, as displayed in Figure \ref{fig:expl_main_financialgoal_context}. Those whose primary financial goal was to pay off debt were among the least comfortable with all interventions. These findings signal the key task of respecting financial goals when designing supportive financial systems.

\begin{figure}[!ht]
  \centering
  \includegraphics[scale=0.5]{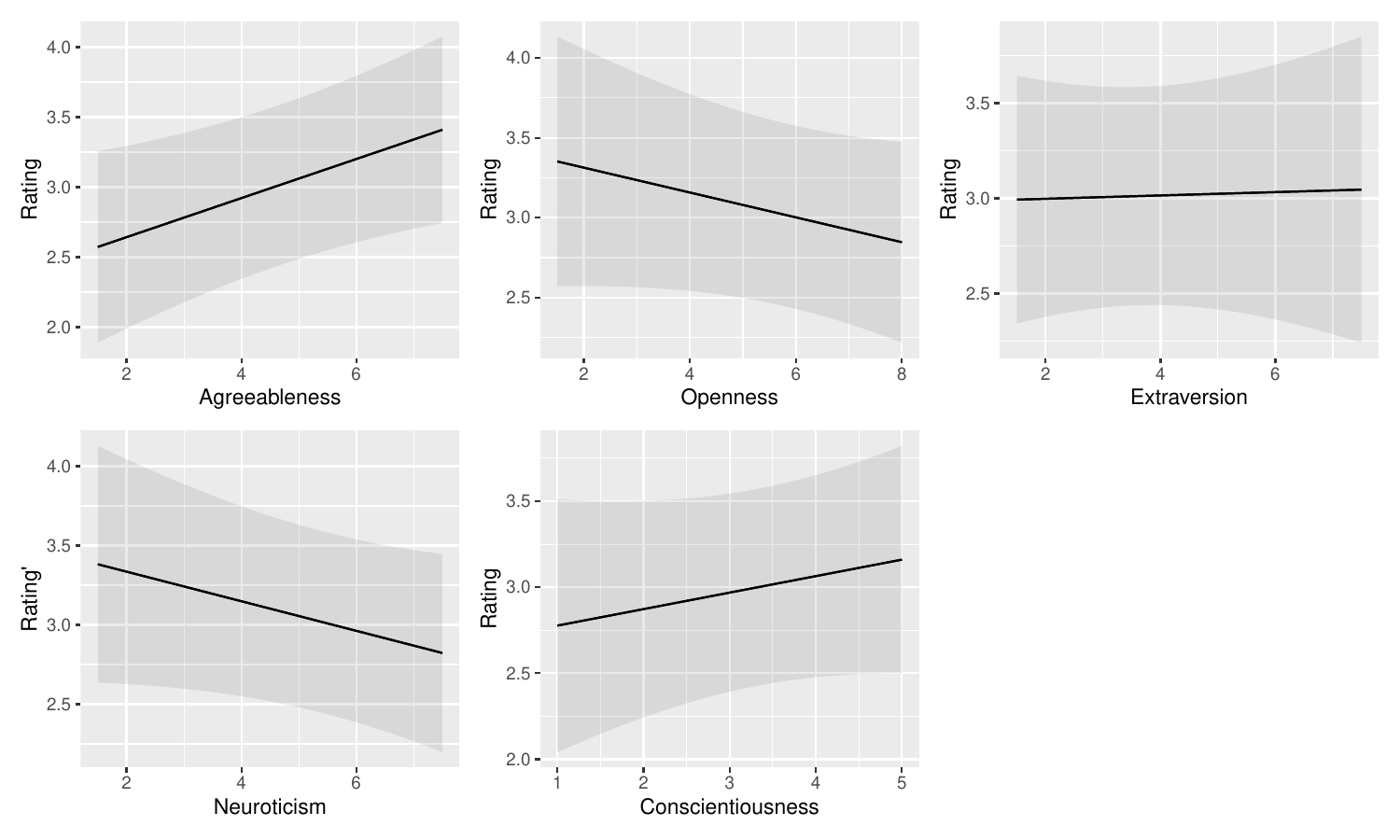}
  \caption{Conditional predictions of comfort ratings per BFI-10 dimension. Agreeableness was significantly associated with increased comfort ratings.}
  \label{fig:expl_bfi10_marginal_predictions}
\end{figure}

Additionally we modeled dimensions of personality and vignette factors to assess the interplay of personality and level of comfort. Average marginal predictions for each BFI-10 dimension are presented in Figure \ref{fig:expl_bfi10_marginal_predictions}. On average, low levels of neuroticism or openness and high levels of extraversion or conscientiousness were associated with higher overall comfort ratings.

\section{Discussion}

We anticipated that participants would be more comfortable sharing data for financial interventions during periods of mood symptoms (H1.1). We interpret this negative finding to represent potential comfort with interventions remaining constant across mood fluctuations. Further, this may reflect the uncertainty surrounding symptom onset: Respondents may prefer continual availability of a given intervention as a protective measure against this uncertainty. 

We also anticipated comfort ratings would be at their highest when sharing spending behaviors during mood symptoms (H1.2). Participants in our sample were instead most comfortable seeking help with planning and budgeting regardless of mood symptoms. We interpret this negative finding to represent an unmet design opportunity for goal-oriented, collaborative financial planning, where people living with BD and their care partners might collaboratively engage with a digital tool to identify and progress towards financial goals. 

Whether an individual had previously sought help from others in managing impulsive spending was associated with significant increases to comfort ratings. This suggests that some individuals are already open about their illness and, potentially, their related financial issues, while others may be intensely private about both. We argue that this distinction in help-seeking preferences may serve as an important indicator of willingness to involve others at all.

\subsection{Conclusion and Future Work}

This study identifies several contributing factors to the acceptability of sharing financial data for third-party illness management. This work extends our prior survey research which identified that individuals with bipolar disorder were quite comfortable sharing financial data for their own financial self-management \cite{brozenasupportive2024}. Held together, these findings suggest that the design of hypothetical, just-in-time digital interventions promoting financial stability (via self-management and, when desired, third-party involvement) would be acceptable to this population.

Supportive financial systems might incorporate just-in-time interventions informed by psychological therapies such as Cognitive Behavioral Financial Therapy \cite{klontzfinancial2015}, DBT (``urge surfing'', in particular) \cite{reinharthEffects}, and Acceptance and Commitment Therapy \cite{el-sayedefficacy2023, kroskaoptimizing2020, wadaacceptance2015}. One particularly interesting therapeutic target could be Emotion Related Impulsivity (ERI). Prior work has investigated the use of ERI-related digital interventions to reduce aggression \cite{johnsondevelopment2020} and rumination \cite{allencognitive2025} in BD. ERI is highly correlated with both risky decision-making and delay discounting \cite{elliottemotionrelated2023} and remains elevated during euthymic states \cite{johnsonEmotionRelated2018}. As such, digital interventions aiming to reduce delay discounting could be deployed during euthymic periods with potential benefits to risky decision-making during those periods. More specifically, an intervention might be designed around episodic future thinking which has been shown to reduce delay discounting \cite{rungexperimental2018, scholtenBehavioral2019} and has been deployed digitally \cite{perssonepisodic2023}.

Given the high respondent ratings surrounding help with planning and budgeting, supportive financial systems could involve care partners in financial management without explicitly sharing decision-making responsibilities \cite{blairknowing2023}. Care partners could be presented with financial data in an aggregated form of the patient's choosing, effectively adding a layer of privacy through abstraction. An individual may wish, for example, to share only the daily frequency of transactions, or whether they had recently made large online purchases late at night, or if they had exceeded specific budget categories over the last month. Aggregation has been found to be highly acceptable when sharing illness-related data with others \cite{hoefermultiplicative2021}.

In closing, we argue that issues of financial stability warrant increased inclusion into polyadic, collaborative care management for individuals living with bipolar disorder who are open to the prospect. Patients, clinicians, and care partners could collaborate in tandem towards financial stability by incorporating financial insights into shared decision-making dialogues (e.g., the Three Talk model \cite{elwynthreetalk2017}). Polyadic financial management systems tailored to the needs of BD might support these dialogues by following cyclical intervention process models appropriate for episodic illness \cite{vandeneijnde-damencollaborative2024}. More specifically, care teams might collaborate to (1) create financial crisis plans to limit future financial harms, (2) maintain vigilance over early warning signs of financial instability, and (3) offer support during periods of euthymic recovery. Future financial systems can support these recovery-oriented tasks by appropriately situating within existing networks of care.

\section{Disclosures}

\subsection{Author contributions}

Overall study aims were developed by authors 1, 2, 5, 6 and 7. Authors 1 and 2 developed the main survey protocol, with the advisement and guidance of authors 3-7. Authors 3, 4, and 5 provided clinical insights throughout the survey development analysis. Author 1 performed the main rounds of survey analysis with further review provided by authors 2 and 7. The manuscript was drafted by author 1, with rounds of revision provided by authors 2 and 7. All authors reviewed and provided feedback on the submitted manuscript.

\subsection{Acknowledgements}

Research reported in this publication was supported in part by the National Institutes of Health's National Institute of Mental Health under award number R21MH131924 and by the National Science Foundation Graduate Research Fellowship Program under grant number DGE1255832. This manuscript is subject to the NIH Public Access Policy. Through acceptance of this federal funding, NIH has been given a right to make this manuscript publicly available in PubMed Central upon the official date of publication, as defined by NIH. The content is solely the responsibility of the authors and does not necessarily represent the official views of the National Institutes of Health or the National Science Foundation.

\subsection{Financial Disclosure and Conflict of Interest}
TR has been paid to deliver training about psychological approaches to bipolar disorder and working with financial difficulties in mental health. He receives royalties on a book about Bipolar disorder and has been paid to develop therapy material on this area. TR receives royalties from ‘Space from Money Worries’ developed by Silvercloud Health, and has shares in ‘Tell’, a company identifying and supporting vulnerable customers, both of which are not related to the current research. The remaining authors declare no potential conflict of interests.

\printbibliography

\newpage
\appendix

\section{Demographics}\label{app-1}
\begin{longtable}{ll}
\hline
\textbf{Characteristic} & \textbf{Freqs (\% of Valid)} \\

\textbf{Gender} & \\
\hspace{1em}Man & 159 (31.9\%) \\
\hspace{1em}Non-binary & 33 (6.6\%) \\
\hspace{1em}Prefer not to disclose & 3 (0.6\%) \\
\hspace{1em}Prefer to self-describe & 5 (1.0\%) \\
\hspace{1em}Woman & 299 (59.9\%) \\

\textbf{Age} & \\
\hspace{1em}18 to 24 & 64 (12.8\%) \\
\hspace{1em}25 to 34 & 186 (37.3\%) \\
\hspace{1em}35 to 44 & 124 (24.8\%) \\
\hspace{1em}45 to 54 & 68 (13.6\%) \\
\hspace{1em}55 to 64 & 38 (7.6\%) \\
\hspace{1em}65 to 74 & 16 (3.2\%) \\
\hspace{1em}75 or older & 3 (0.6\%) \\

\textbf{Education} & \\
\hspace{1em}Master's degree (MS) & 39 (7.8\%) \\
\hspace{1em}Other & 4 (0.8\%) \\
\hspace{1em}Primary school & 6 (1.2\%) \\
\hspace{1em}Professional/doctorate degree (MD/PhD/JD/etc.) & 97 (19.4\%) \\
\hspace{1em}Secondary/high school (or equivalent) & 150 (30.1\%) \\
\hspace{1em}Some university & 136 (27.3\%) \\
\hspace{1em}University graduate (4 year) & 61 (12.2\%) \\
\hspace{1em}Vocational/technical school (2 year) & \\

\textbf{Employment Status} & \\
\hspace{1em}Employed full time (35+ hrs/week) & 206(41.4\%) \\
\hspace{1em}Employed part time (\textless35 hrs/week) & 51 (10.3\%) \\
\hspace{1em}Homemaker/Stay at home parent & 26 (5.2\%) \\
\hspace{1em}Retired & 11 (2.2\%) \\
\hspace{1em}Self-Employed & 44 (8.9\%) \\
\hspace{1em}Student and not working & 23 (4.6\%) \\
\hspace{1em}Unable to work - receiving disability benefits & 46 (9.3\%) \\
\hspace{1em}Unable to work - not receiving disability benefits & 30 (6.0\%) \\
\hspace{1em}Unemployed, looking for work & 51 (10.3\%) \\
\hspace{1em}Unemployed, not looking & 9 (1.8\%) \\

\textbf{Income Level} & \\
\hspace{1em}0 to 19,999 & 204 (40.9\%) \\
\hspace{1em}20,000 to 34,999 & 112 (22.4\%) \\
\hspace{1em}35,000 to 69,999 & 117 (23.4\%) \\
\hspace{1em}70,000 to 99,999 & 41 (8.2\%) \\
\hspace{1em}100,000 or more & 25 (5.0\%) \\

\textbf{Main Financial Goal} & \\
\hspace{1em}To avoid accruing debt & 45 (9.0\%) \\
\hspace{1em}To build savings/emergency fund & 141 (28.3\%) \\
\hspace{1em}To pay off debt & 128 (25.7\%) \\
\hspace{1em}To reduce impulsive spending & 97 (19.4\%) \\
\hspace{1em}To remember to pay bills on time & 36 (7.2\%) \\
\hspace{1em}To work towards a large purchase & 52 (10.4\%) \\

\textbf{Has used Pay Later Services} & \\
\hspace{1em}I do not know & 6 (1.2\%) \\
\hspace{1em}No & 186 (37.3\%) \\
\hspace{1em}Yes & 307 (61.5\%) \\

\textbf{Marital Status} & \\
\hspace{1em}Divorced & 43 (8.6\%) \\
\hspace{1em}In a relationship, not living together & 49 (9.8\%) \\
\hspace{1em}Living with a partner & 94 (18.8\%) \\
\hspace{1em}Married with children & 77 (15.4\%) \\
\hspace{1em}Married without children & 47 (9.4\%) \\
\hspace{1em}Separated & 12 (2.4\%) \\
\hspace{1em}Single, never married & 172 (34.5\%) \\
\hspace{1em}Widowed & 5 (1.0\%) \\

\textbf{Ethnicity} & \\
\hspace{1em}American Indian/Alaska Native & 4 (0.8\%) \\
\hspace{1em}Asian & 11 (2.2\%) \\
\hspace{1em}Black or African American & 44 (8.8\%) \\
\hspace{1em}More than One Race & 46 (9.2\%) \\
\hspace{1em}Unknown or Not Reported & 12 (2.4\%) \\
\hspace{1em}White & 382 (76.6\%) \\

\textbf{Hispanic} & \\
\hspace{1em}No & 440 (88.2\%) \\
\hspace{1em}Yes & 59 (11.8\%) \\

\textbf{Bipolar Subtype} & \\
\hspace{1em}Bipolar Disorder NOS & 119 (23.8\%) \\
\hspace{1em}Cyclothymia & 9 (1.8\%) \\
\hspace{1em}I do not know & 40 (8.0\%) \\
\hspace{1em}Type I & 115 (23.0\%) \\
\hspace{1em}Type II & 216 (43.3\%) \\

\textbf{Age at Diagnosis} & \\
\hspace{1em}Under 12 years & 16 (3.2\%) \\
\hspace{1em}12--18 years & 98 (19.7\%) \\
\hspace{1em}19--29 years & 243 (48.8\%) \\
\hspace{1em}30--39 years & 94 (18.9\%) \\
\hspace{1em}40--49 years & 30 (6.0\%) \\
\hspace{1em}50+ years & 17 (3.4\%) \\

\textbf{Prior Hospitalization} & \\
\hspace{1em}No & 230 (46.1\%) \\
\hspace{1em}Not sure & 16 (3.2\%) \\
\hspace{1em}Yes & 253 (50.7\%) \\

\textbf{Psychiatric Advance Directive} & \\
\hspace{1em}No & 422 (84.6\%) \\
\hspace{1em}Not sure & 37 (7.4\%) \\
\hspace{1em}Yes & 40 (8.0\%) \\

\textbf{Is Banked} & \\
\hspace{1em}I do not know & 1 (0.2\%) \\
\hspace{1em}No & 26 (5.3\%) \\
\hspace{1em}Yes & 468 (94.5\%) \\

\textbf{Has Savings Account} & \\
\hspace{1em}I do not know & 3 (0.6\%) \\
\hspace{1em}No & 135 (27.1\%) \\
\hspace{1em}Yes & 360 (72.3\%) \\

\textbf{Bankruptcy Declared} & \\
\hspace{1em}No & 430 (86.2\%) \\
\hspace{1em}Not sure & 12 (2.4\%) \\
\hspace{1em}Yes & 57 (11.4\%) \\

\textbf{Bankruptcy Considered} & \\
\hspace{1em}No & 326 (65.3\%) \\
\hspace{1em}Not sure & 15 (3.0\%) \\
\hspace{1em}Yes & 158 (31.7\%) \\

\textbf{Requested Financial Help from Care Partner} & \\
\hspace{1em}No & 321 (64.6\%) \\
\hspace{1em}Not sure & 9 (1.8\%) \\
\hspace{1em}Yes & 167 (33.6\%) \\

\textbf{Ease in Trusting Clinicians} & \\
\hspace{1em}Extremely difficult & 35 (7.0\%) \\
\hspace{1em}Extremely easy & 112 (22.5\%) \\
\hspace{1em}Neither easy nor difficult & 62 (12.4\%) \\
\hspace{1em}Somewhat difficult & 78 (15.7\%) \\
\hspace{1em}Somewhat easy & 211 (42.4\%) \\

\textbf{Ease in Trusting Care Partner} & \\
\hspace{1em}Extremely difficult & 21 (4.2\%) \\
\hspace{1em}Extremely easy & 157 (31.5\%) \\
\hspace{1em}Neither easy nor difficult & 88 (17.7\%) \\
\hspace{1em}Somewhat difficult & 52 (10.4\%) \\
\hspace{1em}Somewhat easy & 180 (36.1\%) \\

\textbf{Ease in Trusting Banks} & \\
\hspace{1em}Extremely difficult & 200 (40.2\%) \\
\hspace{1em}Extremely easy & 17 (3.4\%) \\
\hspace{1em}Neither easy nor difficult & 103 (20.7\%) \\
\hspace{1em}Somewhat difficult & 131 (26.3\%) \\
\hspace{1em}Somewhat easy & 47 (9.4\%) \\
\end{longtable}

\newpage

\section{Marginal Predictions}\label{app-2}
\begin{longtable}{lcc}
\hline
\textbf{Characteristic} & \textbf{Average Marginal Predictions} & \textbf{95\% CI}\\

\textbf{Actors} &  & \\
\hspace{1em}Banks & 3.1 & 2.7, 3.5\\
\hspace{1em}Care Partner & 5.0 & 4.6, 5.4\\
\hspace{1em}Clinicians & 4.4 & 4.1, 4.8\\

\textbf{Intervention Context} &  & \\
\hspace{1em}48-hour Spending Restriction & 3.1 & 2.7, 3.5\\
\hspace{1em}Help with Planning \& Budgeting & 5.1 & 4.7, 5.5\\
\hspace{1em}Share Spending Details & 4.4 & 4.0, 4.8\\

\textbf{Mood State} &  & \\
\hspace{1em}Euthymic & 4.0 & 3.6, 4.3\\
\hspace{1em}Symptomatic & 4.4 & 4.1, 4.7\\

\textbf{Age Range} &  & \\
\hspace{1em}18 to 24 & 4.2 & 3.6, 4.7\\
\hspace{1em}25 to 34 & 4.3 & 4.0, 4.7\\
\hspace{1em}35 to 44 & 4.2 & 3.8, 4.6\\
\hspace{1em}45 to 54 & 4.0 & 3.5, 4.5\\
\hspace{1em}55 to 64 & 3.8 & 3.2, 4.5\\
\hspace{1em}65 to 74 & 4.3 & 3.3, 5.3\\
\hspace{1em}75 or older & 3.8 & 1.5, 6.1\\

\textbf{Gender} &  & \\
\hspace{1em}Man & 4.4 & 4.0, 4.7\\
\hspace{1em}Non-binary & 4.9 & 4.2, 5.7\\
\hspace{1em}Prefer not to disclose & 4.6 & 2.4, 6.8\\
\hspace{1em}Prefer to self-describe & 3.9 & 2.2, 5.7\\
\hspace{1em}Woman & 4.0 & 3.7, 4.3\\

\textbf{Education} &  & \\
\hspace{1em}Master's degree (MS) & 4.6 & 3.9, 5.3\\
\hspace{1em}Other & 3.9 & 1.9, 5.9\\
\hspace{1em}Primary school & 3.8 & 2.1, 5.4\\
\hspace{1em}Professional or doctorate degree (MD, PhD, JD, etc.) & 3.7 & 2.1, 5.4\\
\hspace{1em}Secondary/high school (or equivalent) & 4.2 & 3.7, 4.7\\
\hspace{1em}Some university & 4.1 & 3.7, 4.5\\
\hspace{1em}University graduate (4 year) & 4.2 & 3.8, 4.6\\
\hspace{1em}Vocational/technical school (2 year) & 4.2 & 3.7, 4.8\\

\textbf{Income} &  & \\
\hspace{1em}0 to 19 999 & 4.1 & 3.8, 4.4\\
\hspace{1em}100 000 or more & 4.2 & 3.3, 5.1\\
\hspace{1em}20 000 to 34 999 & 4.3 & 3.8, 4.7\\
\hspace{1em}35 000 to 69 999 & 4.3 & 3.8, 4.7\\
\hspace{1em}70 000 to 99 999 & 4.3 & 3.6, 4.9\\

\textbf{Ethnicity} &  & \\
\hspace{1em}American Indian/Alaska Native & 4.9 & 3.0, 6.8\\
\hspace{1em}Asian & 4.7 & 3.5, 5.9\\
\hspace{1em}Black or African American & 4.6 & 3.9, 5.2\\
\hspace{1em}More than One Race & 4.2 & 3.6, 4.8\\
\hspace{1em}Unknown or Not Reported & 4.3 & 3.1, 5.5\\
\hspace{1em}White & 4.1 & 3.8, 4.4\\

\textbf{Hispanic} &  & \\
\hspace{1em}No & 4.2 & 3.9, 4.5\\
\hspace{1em}Yes & 4.1 & 3.5, 4.7\\

\textbf{Diagnosis Subtype} &  & \\
\hspace{1em}Bipolar Disorder Not Otherwise Specified & 4.1 & 3.7, 4.5\\
\hspace{1em}Cyclothymia & 3.4 & 2.1, 4.8\\
\hspace{1em}I do not know & 4.1 & 3.4, 4.8\\
\hspace{1em}Type I & 4.1 & 3.7, 4.5\\
\hspace{1em}Type II & 4.3 & 4.0, 4.7\\

\textbf{Prior Hospitalzation} &  & \\
\hspace{1em}No & 4.0 & 3.7, 4.4\\
\hspace{1em}Not sure & 4.2 & 3.1, 5.3\\
\hspace{1em}Yes & 4.3 & 4.0, 4.7\\

\textbf{Banked} &  & \\
\hspace{1em}I do not know & 5.9 & 1.1, 11\\
\hspace{1em}No & 4.2 & 3.3, 5.1\\
\hspace{1em}Yes & 4.2 & 3.9, 4.5\\

\textbf{Savings} &  & \\
\hspace{1em}I do not know & 4.7 & 1.8, 7.6\\
\hspace{1em}No & 4.3 & 3.9, 4.7\\
\hspace{1em}Yes & 4.1 & 3.9, 4.4\\

\textbf{Ease in Trusting Clinician} &  & \\
\hspace{1em}Extremely difficult & 3.5 & 2.7, 4.3\\
\hspace{1em}Extremely easy & 4.9 & 4.4, 5.3\\
\hspace{1em}Neither easy nor difficult & 3.6 & 3.0, 4.1\\
\hspace{1em}Somewhat difficult & 3.6 & 3.0, 4.1\\
\hspace{1em}Somewhat easy & 4.4 & 4.0, 4.7\\

\textbf{Ease in Trusting Care Partner} &  & \\
\hspace{1em}Extremely difficult & 3.8 & 2.8, 4.9\\
\hspace{1em}Extremely easy & 4.4 & 4.0, 4.8\\
\hspace{1em}Neither easy nor difficult & 4.2 & 3.7, 4.6\\
\hspace{1em}Somewhat difficult & 4.3 & 3.6, 4.9\\
\hspace{1em}Somewhat easy & 4.1 & 3.7, 4.4\\

\textbf{Ease in Trusting Banks} &  & \\
\hspace{1em}Extremely difficult & 3.5 & 3.1, 3.8\\
\hspace{1em}Extremely easy & 5.3 & 4.3, 6.3\\
\hspace{1em}Neither easy nor difficult & 4.4 & 4.0, 4.9\\
\hspace{1em}Somewhat difficult & 4.5 & 4.1, 4.9\\
\hspace{1em}Somewhat easy & 5.4 & 4.7, 6.0\\

\textbf{CFPB Financial Wellbeing Score} &  & \\
\hspace{1em}21 & 3.7 & 2.7, 4.8\\
\hspace{1em}57 & 4.1 & 3.8, 4.4\\
\hspace{1em}63 & 4.2 & 3.9, 4.5\\
\hspace{1em}65 & 4.2 & 4.0, 4.5\\
\hspace{1em}82 & 4.4 & 3.8, 5.0\\

\textbf{Asked for Help with Impulsive Spending} &  & \\
\hspace{1em}No & 3.8 & 3.5, 4.1\\
\hspace{1em}Not sure & 4.6 & 3.3, 5.9\\
\hspace{1em}Yes & 4.9 & 4.5, 5.2\\

\end{longtable}

\end{document}